\documentclass[journal]{IEEEtran}
\usepackage{graphicx}	
\usepackage{setspace}
\usepackage{multirow}
\usepackage{booktabs}
\usepackage{tikz}
\usepackage[normalsize]{caption}
\usepackage{float}
\usepackage[fleqn]{amsmath}
\usepackage{color, colortbl}
\usepackage{xcolor}
\usepackage{subfig}
\usepackage{epstopdf}
\usepackage{lipsum}
\usepackage{dblfloatfix}
\usepackage{algorithm2e}
\definecolor{Gray}{gray}{0.9}

\begin{document}
\title{A Diversity-based Substation Cyber Defense Strategy utilizing Coloring Games }

\author{Md~Touhiduzzaman,~and~Adam~Hahn,~\IEEEmembership{Member,~IEEE,}
        and~Anurag~Srivastava,~\IEEEmembership{Senior~Member,~IEEE}}
        
\maketitle

\begin{abstract}

Growing cybersecurity risks in the power grid require that utilities implement a variety of security mechanism $(SM)$ composed mostly of VPNs, firewalls, or other custom security components. While they provide some protection, they might contain software vulnerabilities which can lead to a cyber-attack. In this paper, the severity of a cyber-attack has been decreased by employing a diverse set of $SM$ that reduce repetition of a single vulnerability. This paper focuses on the allocation of diverse \textit{SM} and tries to increase the security of the cyber assets located within the electronic security perimeter(ESP) of a substation. 
We have used a graph-based coloring game in a distributed manner to allocate diverse \textit{SM} for protecting the cyber assets. The vulnerability assessment for power grid network is also analyzed using this game theoretic method. An improved, diversified \textit{SM}s for worst-case scenario has been demonstrated by reaching the Nash equilibrium of graph coloring game. As a case study, we analyze the IEEE-14  and IEEE-118 bus system, observe the different distributed coloring algorithm for allocating diverse \textit{SM} and calculating the overall network criticality. 
\end{abstract}


\section{Introduction}
\IEEEPARstart{O}{ver} the past 15 years, the North American bulk power system has become more prone to the risk of coordinated High Impact Low Frequency (HILF) cyber attack due to growing dependency on digital communicating equipment for substation automation~\cite{energy}. Concerns for the cybersecurity of the power network has increased since December $23^{rd}, 2015$ when an attacker successfully intruded a Ukrainian substation, tripped the substation circuit breaker. This resulted in a substantial blackout~\cite{ukrain}. Nowadays, software vulnerabilities have become major a concern for power grid network. 
All public known vulnerabilities are listed in common vulnerabilities and exposure (CVE) list which require extensive analysis for risk management process. Recent trend  analysis shows that more than $80\%$ of total vulnerabilities are exploitable by network access control~\cite{president}, hence there is a need for increased for security mechanism standards.

The North American Electric Reliability Corporation (NERC) has introduced the Critical Infrastructure Protection (CIP) standards to protect the bulk-power system from cyber-attack. NERC standards include the ESP which is used to prevent remote intrusion to the sensitive internal system, and the substation residing within this perimeter. According to ESP, each substation is to be equipped with a set of security control mechanisms based on their criticality. 
Network security risks such as software exploitation exist in substation automation due to the lack of security feature (e.g., confidentiality, authentication,etc.) in communication layer.

It is well documented that having diversity on critical systems is an important aspect of improving the overall security. The same idea is extended in the cybersecurity domain where diversity in the software platforms on a  security mechanisms prevent single point of failure scenarios. However, no research work has been done in analyzing diversity for grid security. Without diversity, a single exploited vulnerability on software exploitation can provide access to multiple substations. Unfortunately, the power grid shows a very high level of homogeneity where each substation relies on the limited number of vendors to build their security infrastructure. It is thus possible for an attacker to travel across the entire power network and reduce the system-level robustness given by traditional planning and operational criteria. This propagation behavior of a cyber-attack can be minimized by utilizing diverse \textit{SM}. In this manner, an attack requires more exploits/resources.

The game theoretical approaches are used for modeling and analyzing network behavior across the network where players compete for finite resources \cite{game}. In our paper, network security heterogeneity has been achieved by using a combinatorial optimization polymatrix graph coloring game. Work by Chaudhuri~\cite{chaudhuri} first introduced the theoretical background of a network coloring game. Based on Chaudhuri, we propose a graph coloring game that assigns a limited number of software packages based on their security strength (Action: color) to a set of \textit{SM} (Player: node) under some constraints (Strategies), such as there exists an increase in the security (Payoff: security index) of the cyber assets to the entire power grid network. The main contributions of this paper are summarized as follows:
\vspace{.2em}
\begin{enumerate}
\item Introduce a graph based security model (section IV) where the diversity of SM is achieved by using graph coloring game (section V). \item The security index of each SM is designed by trading off between vulnerabilities of the substation and security strength of that mechanism.
\item A graph coloring game is proposed to identify an optimal software package allocation decision that ensures the highest level security and reduce the attack propagation of overall power grid network. 
\end{enumerate}
\vspace{-.5em}

\section{Related Work and Overview of Cyber Protection on the Power Grid}
\subsection{Related work}
Multiple new metrics have been proposed to determine the security risks of a power grid system \cite{zonouz}\cite{anurag}. A work that analyzes common vulnerability scoring system (CVSS) metric against actual attack in the controlled environment is proposed in ~\cite{actual}. However, none of those metrics considered the diversity of SM on the power grid by considering the defensive strategies.

The use of diversity on $SM$ has gained much attention as an important security property~\cite{gain}. Diversity on $SM$ deployment strategies for resilience has been evaluated ~\cite{resili} and has been found to improve the robustness~\cite{k-zero} of the network against zero-day attack by introducing a network security metric. 
%
Previously, multiple studies have been performed that study survivability through heterogeneity. Source code modification~\cite{software}~\cite{software1} had been proposed to diversify the software packages on computer systems. Keromytis and Prevelakis~\cite{monoculture} modified the environment and structure of the network to achieved the diversity against system monocultures. 

In~\cite{donnell}, the authors proposed a distributed graph coloring algorithm which leverages a malicious node to attack the same software packages, this resulted in software diversity. This work focuses on topological properties of the computer network which is similar to the concepts~\cite{kearns} of preventing human behavior epidemics on social relations.

Recently, game theory has been applied to the distributed algorithm to achieve the proper allocation of resources in cloud computing~\cite{cloud}, peer-to-peer system~\cite{peer} and web cache~\cite{web}. Papagopoulou and Spirakas ~\cite{papa} proposed theoretical background of efficient graph coloring game which was based on local search. 
In~\cite{vertex}, the authors proposed a game-theoretic approach of vertex coloring in a distributed manner for evaluating the performance of the wireless network in a simulated environment.  

In this paper, Our work emphasizes on interdependency, the complex network where a system-wide study of diversity has not yet performed. We focused on the heterogeneity of SM in the substation to reduce the propagation of computer malware. 

\subsection{Overview of cyber protection}The Cyber assets in power system always try to maintain some level of protection strategies; there are remaining questions of how to diversify the set of SM that most accurately reflect the grid's risk. There exists some challenges to achieve a strong defense mechanism for the substation against a cyber-attack. Those challenges are include the management of security keys, poor authentication, and authorization mechanism, fragile legacy devices and unpatched systems. It is mandatory to defend the substation by hardening the interior of operation network and also harden the field sites and their partner connections. By hardening, we are able to limit the dispersal of single point vulnerabilities and diminish the attacker capability to expand a compromise the entire system. 

As an example, In NERC, all the critical cyber assets require that all ESP substations that have been classified as either high/medium or low to provide isolation between untrusted network and substation. NERC CIP-005-5 standard addresses identification and protection of all electronic access point on ESP~\cite{ESP}. The ESP depends on security mechanisms and protected by an electronic access point (EAP) that allows routable communication between cyber assets. According to NERC CIP-005-05, high/medium ESP substations required additional security requirement such as multi-factor authentication and encryption to protect the remote interactive sessions. Figure 1 provides an overview of the required protection strategies in both ESP and LESP strategies, demonstrating the SMs to protect both interactive and SCADA communication sessions.
\vspace{-1em}
\begin{figure}[!ht]
  \centering
  \includegraphics[width=50mm,scale=0.50]{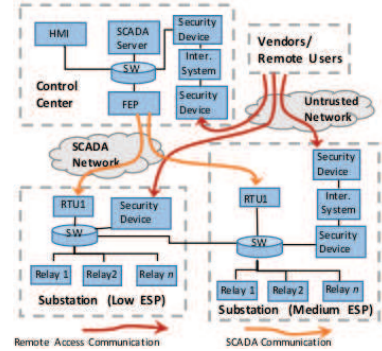}
  \caption{Example substation protection architecture}
  \vspace{-1.5em}
  \label{cip_esp}
\end{figure}


\section{Cyber-Physical Framework }
In this section, we propose the cyber-physical framework where the game-theory is applied to achieve the diversity of SM related to protection of cyber assets at substation. This cyber-physical framework is modeled as a graphical representation named $Security\ Graph, M$. It considers all possible attack paths that an attacker could use to access and manipulate the substation. We explore various distributed algorithms in $Diversity\ Graph, G$. $G$ is extracted from $M$ that captures only SMs installed for substation protection. Fig 2 shows the proposed approach to achieve diversity on SM. 
\begin{figure}[!ht]
  \centering
  \includegraphics[width=80mm,scale=0.80]{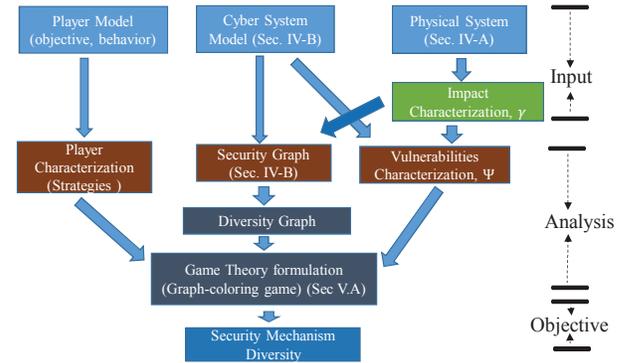}
  \caption{Modeling of achieve diversity using game theory}
   \label{cip_esp}
   \vspace{-1em}
\end{figure}
\vspace{-1em}
\subsection{Physical system criticality}
We assume an intruder attempt to seek strategies to find out the most critical substations and tries to manipulate those substations control parameter to cause damage as much as possible. Hence, from power utility perspective, the most critical substation need to be identified and equipped with well protection devices to protect from cyberattack.
In our proposed method, we categorized the substations into high impact substation(HIS) and low impact substation(LIS) based on their criticality. 
To achieve this, we have used the impact factor calculation~\cite{impact_equation}. The IEEE common data (e.g.bus data, branch data) format is applied to calculate impact factor.
\begin{figure}[!ht]
  \centering
  \includegraphics[width=90mm,height=5.0cm,scale=0.80]{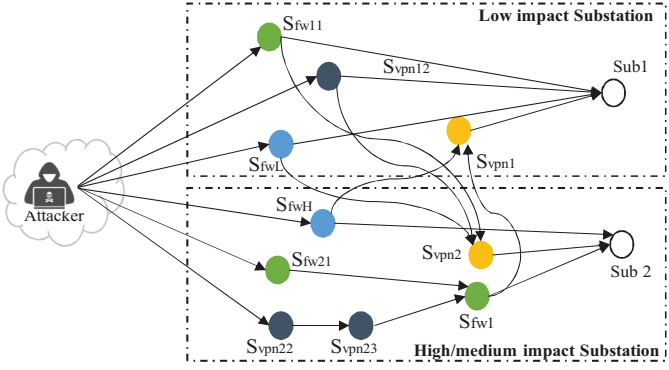}
  \caption{Security Graph Model}
   \label{cip_esp}
   \vspace{-2em}
\end{figure}


The author in ~\cite{impact_equation} introduced the impact factor metrics, $\gamma$ which applies to the analysis of the cyber attack on substations. This impact factor represents the impact of the removal of a single-, double- or multiple- substation from the entire power system by switching it off due to a cyber attack. This impact factor is defined as follows:
\begin{equation}
\gamma= \bigg(\dfrac{P_{lol}}{P_{total}}\bigg)^{L^*-1}
\end{equation}
In this equation, $L^*$ represents the maximum loading level value, where the power flow study diverges. This loading level, $L$ is achieved by performing the continuation power flow methods (i.e., $P-V$ curve analysis). Here,  $P_{lol}$ and $P_{total}$  represents the loss of load and total system load respectively. In this method, substations are designated as the highest level of criticality whose impact factor, $\gamma=1$ and designated as critical if their impact factor is greater than $threshold$. 
System planners have their own impact level threshold based on their security level responsiveness and willingness to invest. If the substation impact factor $\gamma$ is more than this $threshold$ level, then this substation is classifies as HIS.
\vspace{-1em}
\subsection{Cyber model}
The cyber system modeled as a security graph model, $M = (C, K)$ where $C$ is the set of cyber assets and $k$ is the networking link connecting them. The cyber assets include, the $\textit{SM}$s, the substation protection equipment (e.g., circuit breaker, relay), and the attacker. We define a set of $\textit{SM}$s, $S\in  \{VPN, encryption, authentication, firewall\}$ used to protect protection equipment and connected networks $K$. $M$ is developed with the following principles: (i) some security mechanisms are used to protect SCADA communication, and (ii) that multiple SMs could be implemented in a single device, and (iii) that substations are interconnected, i.e. to support transfer trip relay messages between connected substations, and (iv) that other substation devices (e.g., RTUs, relays) do not implement any SM. An example $M$ is presented in Fig. 3, it is modeled based on the substation protection architecture shown in Fig. 1 where HIS and LIS are both connected to each other through VPN. A example set of \textit{SMs} for $M$ can be outlined as follows:

\begin{enumerate}
\item SCADA firewall ($S_{fwH}$, $S_{fwL}$)
\item VPN ($S_{vpn1}$, $S_{vpn2}$)
\item System firewall (Local) ($S_{fw11}$, $S_{fw21}$, $S_{fw1}$)
\item System Authentication ($S_{vpn22}$, $S_{vpn12}$, $S_{vpn23}$)
\end{enumerate}

From this $M$, we need to extract $G$ for further analysis. $G$ is based on the connection of the $\textit{SM}$s. As we mentioned before our goal is to diverse the $SM$. Hence, In this work, we focus only on the vertices which represent the \textit{SM}. 
\vspace{-1em}
\subsection{Threat model}
We need to construct our power grid network by utilizing a diverse set of \textit{SMs} so that a malware will not propagate across the entire network by preventing single point of failure. We developed our threat model by making the following assumption:
\begin{enumerate}
\item A threat is modeled against  $k$ zero-day attacks proposed by~\cite{resili} where $k$ is the number of unique vulnerabilities.
\item A software vulnerability exists such that it compromises all the devices where this software is installed .
\item The protected system should have diversity $x$ greater or equal than the attacker's capability to attack $n$ security mechanism located in attack path $p$.

$ Diversity, x=  \#colors(p)$

$\qquad \qquad s.t.\ \exists  p|p \in G $

$\qquad \qquad \qquad  \mid k \mid < x\leq \mid n \mid \qquad \forall p\in G$
\end{enumerate}

Consider, a simple scenario where the attacker is able to move from one node to another node by using the network. First, consider a case where each type of SMs are running the same software package (i.e.color) (Fig 3). In this case, the attacker can easily compromise substation 2 by exploiting $S_{fw1}$ and $S_{vpn22}$ as their neighbor SM is running the same set of software packages. It clearly indicates a lack of diversity and a need for replacing this software arrangement to prevent malware from propagating to other network systems all at once. As a mitigation proposal, we have installed a different set of software packages so that neighbor node not running the same software packages. In this case, we have alloted green, yellow and black color to represent $S_{vpn23}$, $S_{fw1}$, and $S_{vpn1}$, respectively. Now, let us assume that an attacker wants to access either substation 1 or substation 2 or both by exploiting $k-$ different types of software packages. Table I shows the number of exploited software(k), and feasible attack path to access the substation. By analyzing this penetration problem, we conclude that the diversity will create difficulties for an attacker to attack on the cyber assets across the entire network by reducing the number of attack path.
\begin{table}
\begin{footnotesize}
\captionsetup{aboveskip=5pt, belowskip=5pt}
  \caption{Attack path analysis}
  \label{tbl:excel-table}
  \includegraphics[width=\linewidth]{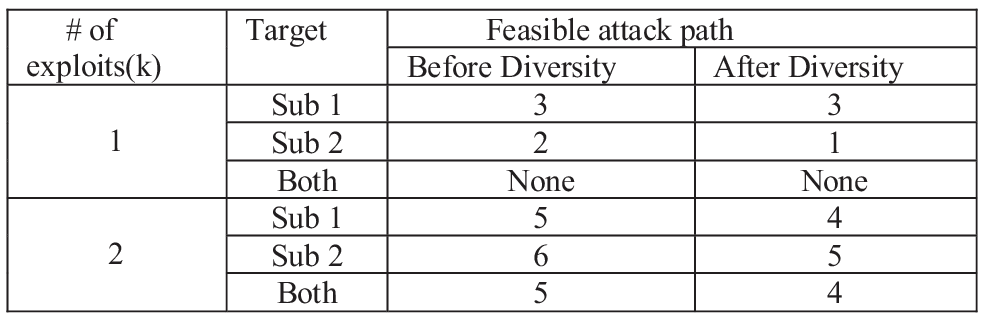}
  \vspace{-3em}
  \end{footnotesize}
\end{table}


\section{Distributed Algorithm}
We applied different distributed coloring algorithms in our diversity graph to achieve diversification of the SM. The goal is to allocate software packages to the SM in such a way that neighboring node should not run the same software package. Each of the software packages are represented by color and associated with an integer value based on security strength variable, where a higher integer value is regarded as being a highly secured software allocated to the \textit{SM}.

In the distributed coloring algorithm, let, G = (V, E) be a finite, undirected diversity graph with $|V | = n$ vertices. Where, $N(v):= \{ w \in V;(v,w)\in E \}$ denotes the set of neighbor nodes for $v\in V$. Each vertex has a set of $x$ colors that represents actions $[ X ] = \{1, . . ., x\}$. The algorithm goal is to choose a profile $c=(c_v)_{v\in V} \in X^n$ from the combination of actions in set $X$, where $c$ is an integer value of color chosen by vertex. The least number of colors required for coloring the entire graph is referred as chromatic number $(\chi)$.

\subsection{Graph Coloring Game}

In this paper, we propose a graph coloring game where each vertex $v$ in $G$ acts as a player who needs to the choose a color according to different strategies. A player payoff is defined as the security index $U^v(c)$ which evaluates the vulnerability. The overall game is played in rounds where each player chooses a color in each round according to their strategies and by observing the colors chosen by neighbors. If a player is able to choose a color different from the colors used by its neighbors players, then it is  \textit{Satisfied}; otherwise, it is \textit{Unsatisfied}. If the player reaches an unsatisfied state  then it most choose another color such as that it becomes satisfied. This processes is repeated until all the players become \textit{Satisfied}, then our graph coloring game reaches its Nash equilibrium. In the next paragraph, a set of more formal rules are given:

The graph coloring game $\Gamma(G)$ is a game of strategic form where the set of vertices $V$ refers as set of the players, and  Each of the player $v\in V$ needs to choose a pure strategy profile $c$ from action set $X$ based on his strategies. Assume, $p$ denotes the type of security mechanism (e.g.VPN, firewall, etc.)

The payoff of a vertex in our game depends on the security index that is defined as,
\begin{equation}
$$ U^{v}(c)=\sum_{w\in N(v)}  |c(v)*\Psi(v)-c(w)*\Psi(w)|$$
\end{equation}
where, $\Psi$ is the vulnerability of the SM. For a set of SM located in an substation $Z$, the security mechanism vulnerability $\Psi(v)$ of SM $v$ is referred  as potential damage over that substation, $Z$.
\begin{equation}
\Psi(v)=\pi^p_{v}\times\gamma(Z)
\end{equation}
where $\pi^p_v$ is the likelihood that a substation is attacked through a specific security mechanism $p$. This security index $U^{v}(c)$ measures the complexity of the cyber attack that is required to exploit the vulnerability once an attacker has gained access to the target security mechanism $v$. 

The above-mentioned security index, $U^v(c)$ identifies the critical SM by considering both the physical impact (vulnerability index) and the difficulty of cyberattack(security strength). For example, if an SM has the same vulnerability index as the neighborhood SM, then $U^{v}$ depends on the difference of the security strength between SMs. The lower the $U^{v}$ of node $v$, the higher the vulnerability of that security mechanism. In this game, the player $v$ needs to choose an appropriate security mechanism to maximize its security index given by (2).

The normalization in (2) put $U^v$ into the same level ((0,10) range) which improve indices integrity and makes it convenient for the further criticality analysis over different electric power system.
\begin{equation}
U^v_{[0,10]}=\dfrac{U^v_i-U^v_{min}}{U^v_{max}-U^v_{min}} 
\end{equation}

Also, we calculate a cumulative security index, $\sigma = \sum_{i=1}^{n} U^{i}(c)$ which indicates how secure system is by determining the diversity of the graph. The higher the $\sigma$, the less critical the components are to the power grid network.

Proper coloring in our game results in the pure Nash equilibrium. Our coloring game reaches Nash equilibrium when all the SM successfully allocates the software packages based on their strategies. In this Nash equilibrium, no player should change their payoff by unilateral deviating. \\
\textbf{Definition 1.} Our security mechanism allocation $X^*$ is said to be pure Nash equilibrium if $U_{v}(X_v^*,X_{-v}^*)\leq U_{v}(X_v,X_{-v}^*)$, $\forall v  \in  n, \forall X_v  \in  X$. Here $X_{-v}^*$ refers the software allocated of all the player except that $v$th vertices.\\
\textbf{Definition 2.} Every pure Nash equilibrium is a proper coloring of graph G.\\
\textbf{Definition 3.}~\cite{definition} For every player $v$ and $c_v,c_{v}^\prime \in X $ and any $c_{-v}$ their exists a generalized ordinal potential function $\phi(.)$ which we have,
\begin{equation}
\begin{aligned}
U^{v} (c_v,c_{-v})-U^{v}(c_{v}^\prime,c_{-v})>0\\
\Rightarrow \phi (c_v,c_{-v})-\phi(c_{v}^\prime,c_{-v})>0
\end{aligned}
\end{equation}

This generalized ordinal potential function admits that our graph coloring game has at least one pure strategy Nash equilibrium~\cite{potential}.

Our graph coloring game is developed based on some strategies. All the strategies play an important role in making our game more solvable and meaningful. These strategies set the rules for a player on how to play the game. All the strategies for game are given below :

$\bullet$ \textit{Bound on the number of colors:} We have a limited number of available software packages i.e. colors. The maximum possible colors available for the game is $\Delta_2(G)+2$. Here, $\Delta$ is the degree of a vertices.
\begin{equation}
x \leq (\Delta_{2}(G)+2)
\end{equation}
 \textbf{lemma 1.}, The total number of colors $x$ satisfies $x\leq\Delta_2(G)+2$ for any pure Nash equilibrium of $\Gamma(G)$ and hence $x\leq\Delta(G)+2$.\\
\textit{Proof:} Let us consider, $x$ is the total number of colors required to achieve a pure Nash equilibrium $c$ of $\Gamma(G)$. If $x=1$, then graph $G$ is disconnected and therefor $\Delta(G)=\Delta_{2}(G)=0$. Now assume, three colors $x_i, x_j, x_k\in X$ are assigned to the graph to color minimum number of vertices. According to Def.2, assume that $n_{x_i}(c)\geq n_{x_j}(c)\geq n_{x_k}\geq n_{x}(c)$ for all colors $x\not\in \{x_i, x_j, x_k\} $ used in proper coloring $c$. Let, the vertex $v$  and her neighbors $w$ assigned the color $x_i$ and $x_j$, respectively. The payoff of that vertex $v$ is $U^v(c)=|x_{i}(v)*\Psi(v)-x_{j}(w)*\Psi(w)|$. Let us assume that there is no edge between $v$ and $w$  with $c_w=x_k$. Then according to Nash equilibrium, $v$ must hold that $n_{x_i}(c)\geq n_{x}(c)+2$. So, the degree of vertex $v$ is the total number of color minus 2, i.e. $\Delta(v)\geq x-2$.

$\bullet$ \textit{Ordering sequence:} We ordered our vertices $v$ by considering the worst-case scenario and this scenario is achieved by choosing a vertex $v$  with maximum criticality, then order the remaining vertices. Figure 5 shows the algorithm for ordering sequence strategy. In this algorithm, $M$ denotes the set of the security mechanism types, $p_i$. Each of the $p_i\in M$ had $v$ number of security mechanisms located on different substations. To make the ordering sequence more feasible, we have considered the degree of each SM.
\begin{figure}[!ht]
\vspace{-1em}
  \centering
    \includegraphics[width=70mm,scale=0.70]{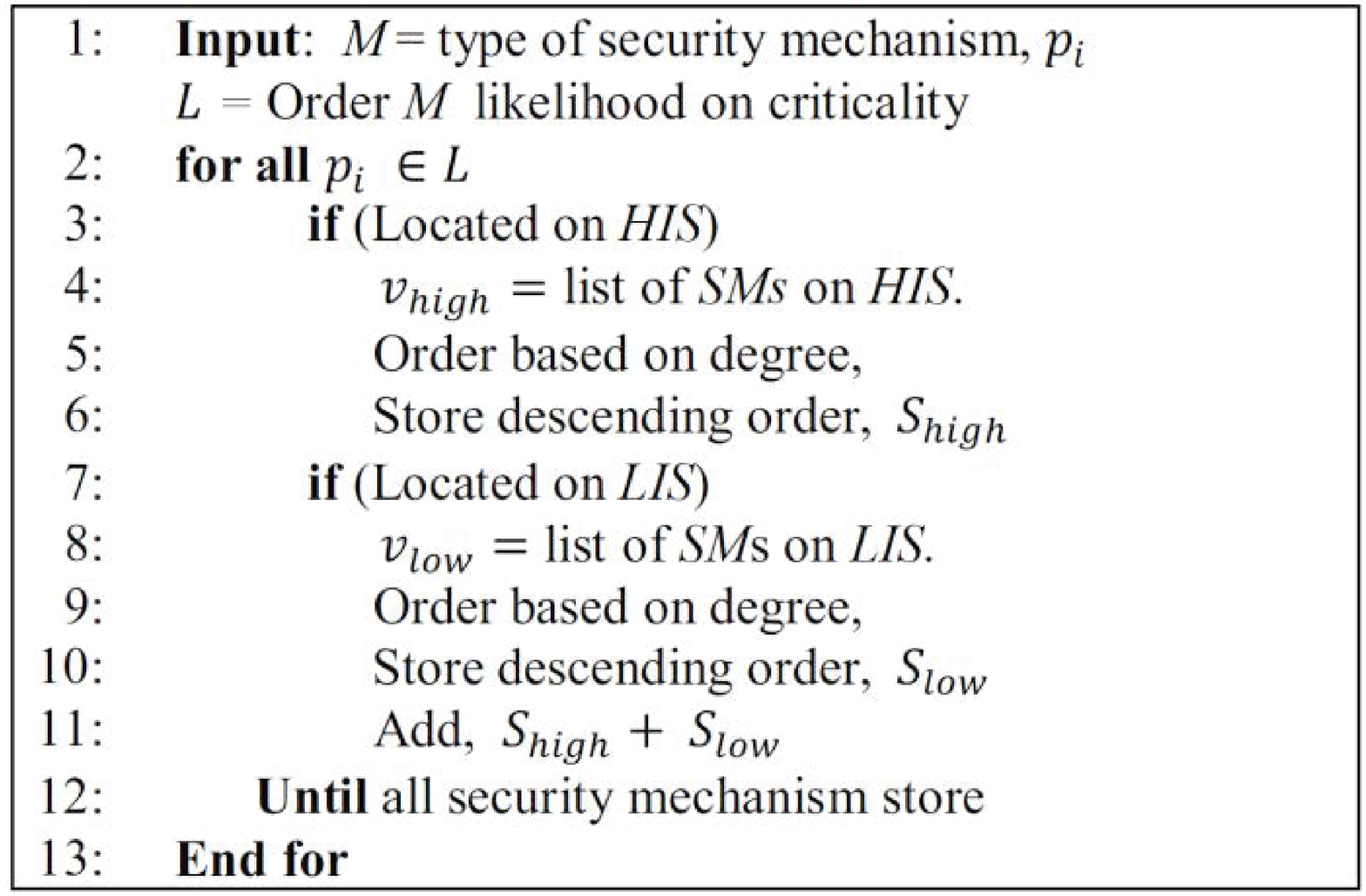}
  \caption{Ordering sequence strategy }
  \label{cip_esp}
\end{figure}

$\bullet$ \textit{Coloring sequence:} Each vertex $v$ needs to choose a strategy profile $c$ from a set of $x$ colors. We consider that the strategy profile $c$ for a specific color is represented by an integer located between $1-10$. As we mentioned before that the higher value of $c$ represents a highly secured software allocated to the \textit{SM}. Our coloring sequence strategy is set in such a way that all important \textit{SM}s found in the ordering sequence strategy can get a higher priority color. This is equivalent to say that the most critical substations have the most secure software combinations.

\subsection{Comparison with non-strategic distributed algorithms }
We have explored different distributed coloring algorithms that try to efficiently allocate \textit{SM}s on the diversity graph. Based on the results there exists some strategical differences between our proposed graph coloring game and other distributed coloring algorithms. The main difference with other distributed coloring algorithms is that they did not consider any worst-case scenario and the physical repercussions of their coloring schemes. 
\subsubsection{Randomized coloring}
In this algorithm, each node $v$ randomly chooses a color from a given list of colors. The number of given color for each node is $d(v)+1$ where, $d(v)$ is the degree of node $v$. This algorithm proceeds in certain rounds and each round, every node randomly picks a color from their given list. Then, they check whether their neighbors pick the same color or not. Any conflict-free node keeps its colors and halt. A node with conflicts withdraw their color, remove that color from their list and continue.
 During the execution of randomized algorithm, all vertices terminated within $O(logn)$ rounds.
\color{black} Figure 6 demonstrates the randomized coloring algorithm. 
\begin{figure}[!ht]
\vspace{-1em}
  \centering
  \includegraphics[width=70mm,scale=0.70]{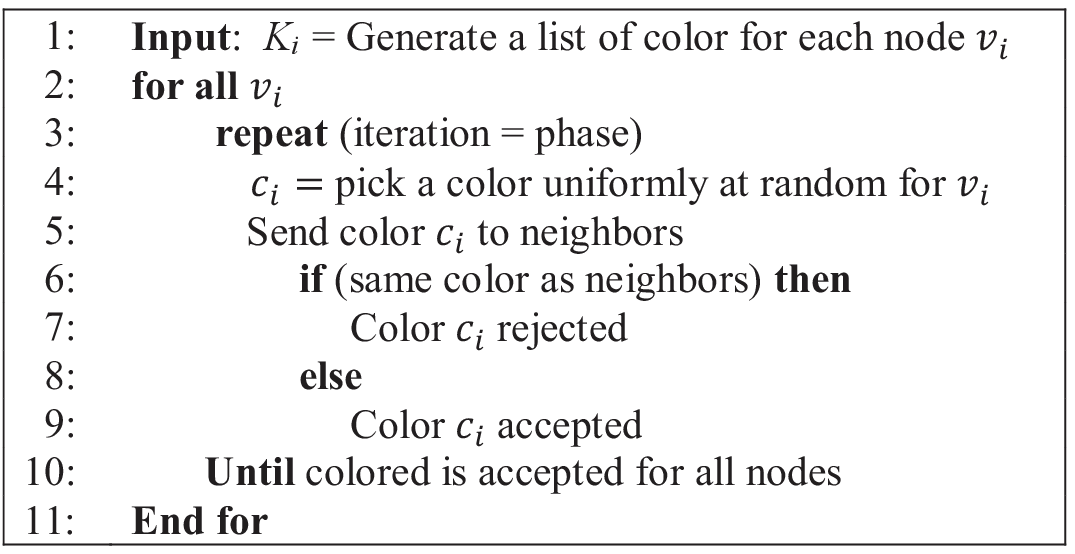}
  \caption{Randomized coloring algorithm}
  \label{cip_esp}
\vspace{-1em}
\end{figure}
\subsubsection{Generic greedy coloring}
In this algorithm, we color the vertices of the graph based on the order of degree. We consider the degree (connectivity) as fundamental property to guarantee the resiliency of a network~\cite{resilience}. In this algorithm, we ordered our SMs according to the descending order of degree. This greedy algorithm is used to find the upper bound of the chromatic number by using Brook’s theorem. This algorithm states that if we order the vertices in descending order based on their degree $(d)$, then chromatic number is, $\zeta = d + 1$. The time complexity of this algorithm increases $O(n^2)$ in each round. Figure 7 shows the generic greedy coloring algorithm.

\begin{figure}[!ht]
\vspace{-1em}
  \centering
  \includegraphics[width=70mm,scale=0.70]{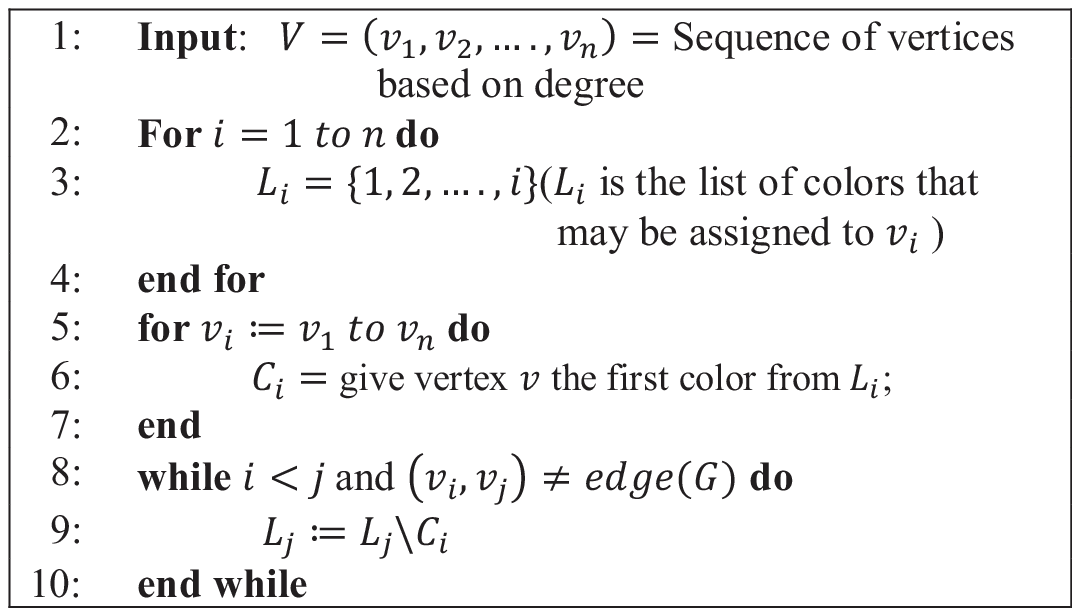}
    \caption{Generic greedy coloring algorithm}
         \label{cip_esp}
\end{figure}

\subsubsection{Sequential coloring}
A sequential coloring algorithm of graph G operating in the following two stages: $(i)$ Determine a coloring sequence $K=(v_1, v_2,....., v_n)$ of vertices in $G$ according to the order of the substation and $(ii)$ pick a color randomly from a list of colors and check whether the neighbor nodes have same color or not. The time complexity of this algorithm is $O(1)$ in each round. 

\section{Simulation Result}
In this paper, the IEEE-14 bus and IEEE-118 bus test case system has been used to evaluate our proposed graph coloring algorithm vs non-strategic distributed algorithms. But mostly our result focus on the analysis of IEEE-14 bus system.

To model our diversity graph $G$, first, we need to develop the security graph $M$ on cyber-physical topology. To do this, we need to identify the most critical substations by performing the impact factor calculation. This impact factor metrics is achieved by performing continuation power flow under normal operating condition. 
Table II shows the impact factor calculation of the IEEE-14 bus system with $\gamma=0.25$ as a threshold value to differentiate between \textit{HIS} and \textit{LIS}.  
The list of HISs and LISs for IEEE-14 bus system are:
$$
Subs_{high}= (2, 3, 4) \quad \textrm{and} \quad
Subs_{low}= (1, 5, 6, 7, 8, 9, 10) $$

\begin{figure*} [ht!]
\centering
    \subfloat[ Randomized coloring algorithm ]{%
       \includegraphics[width=8.2cm,height=4.9cm] {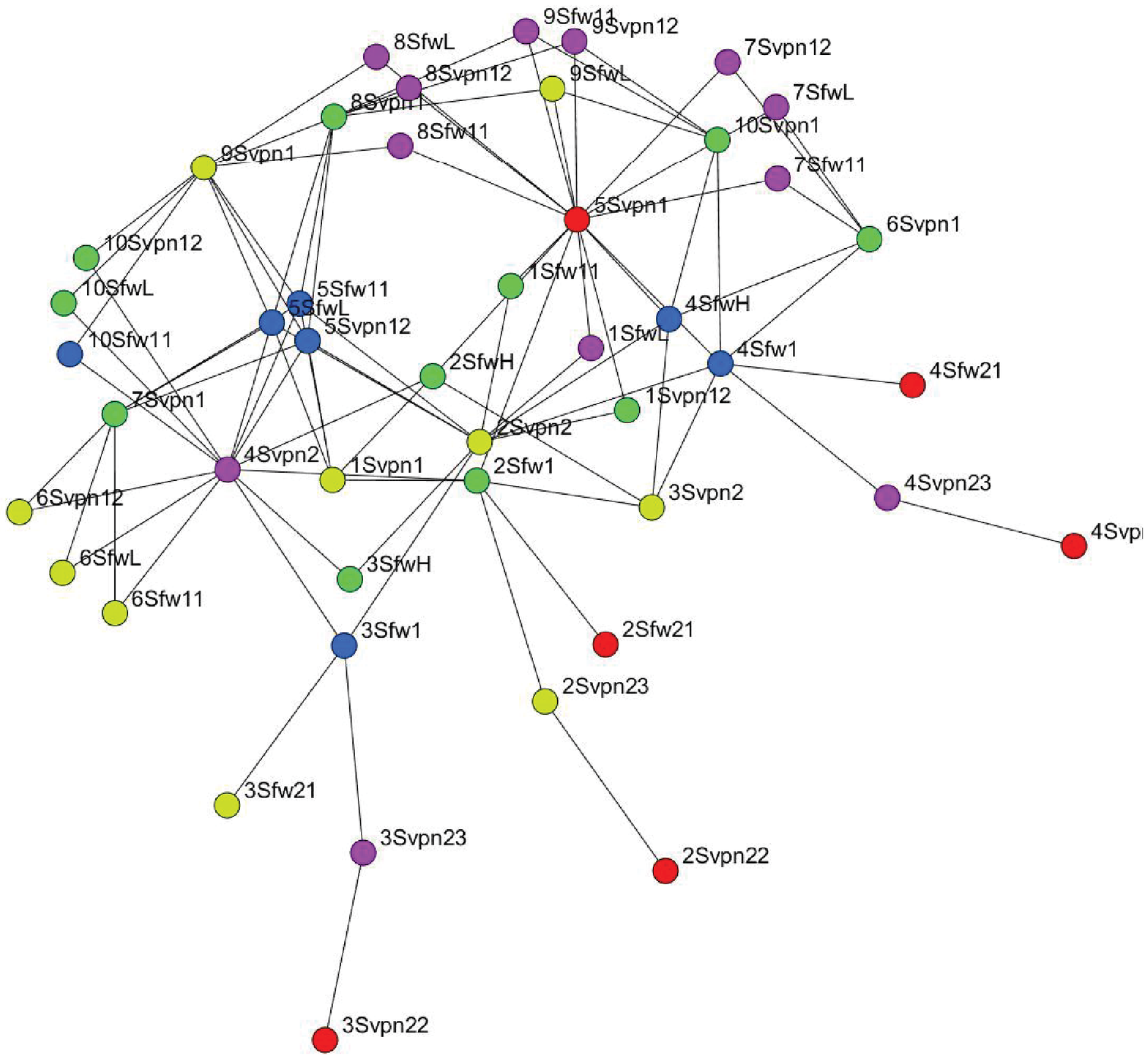}}
    \label{For IEEE-14 bus system}\hfill
  \subfloat[ Greedy coloring algorithm]{%
        \includegraphics[width=8.2cm,height=4.9cm]{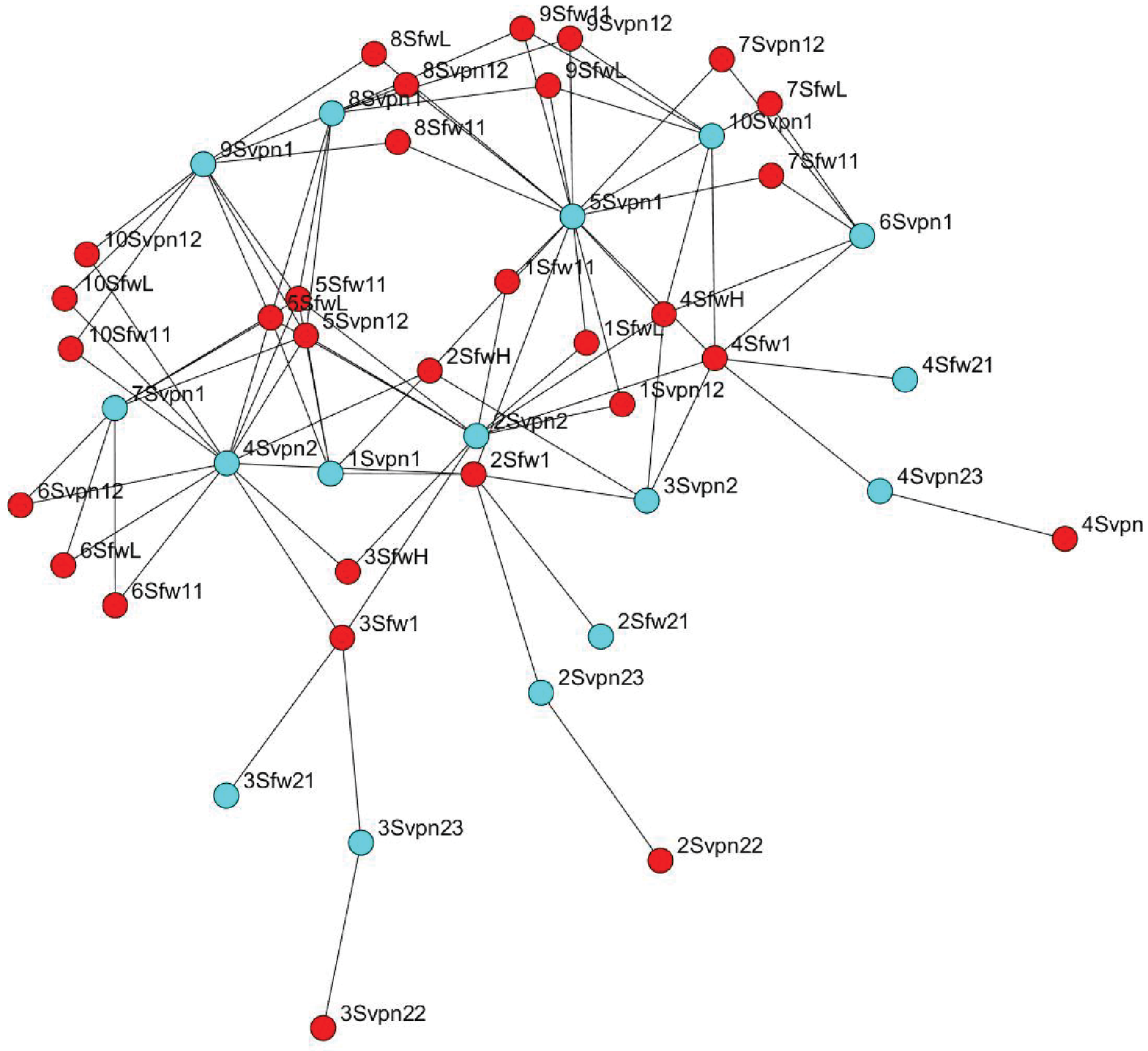}}
    \label{For IEEE-30 bus system}\hfill
    \subfloat[ Sequential coloring algorithm]{%
       \includegraphics[width=8.2cm,height=4.9cm] {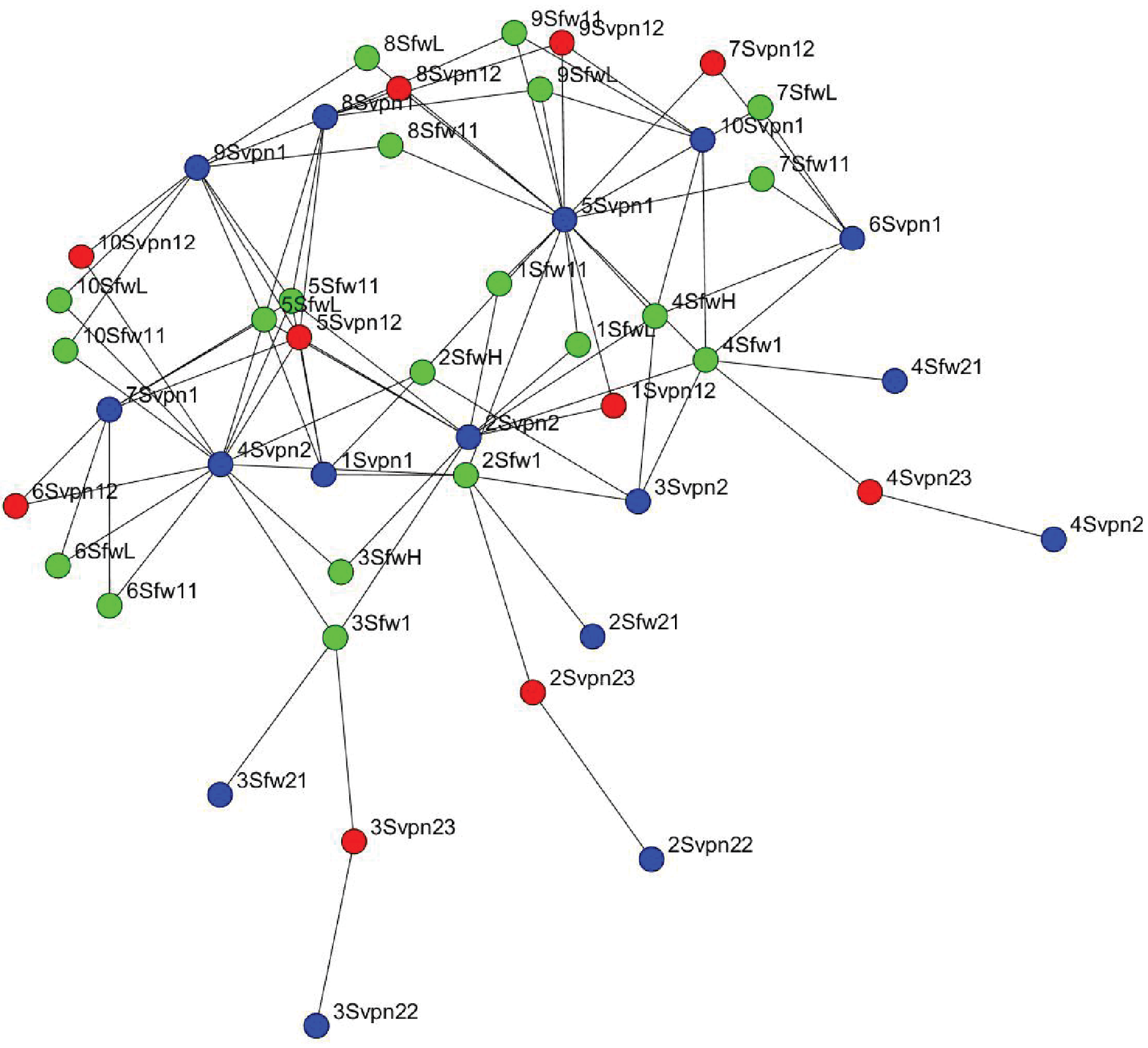}}
    \label{For IEEE-14 bus system}\hfill
  \subfloat[Graph coloring game]{%
        \includegraphics[width=8.2cm,height=4.9cm]{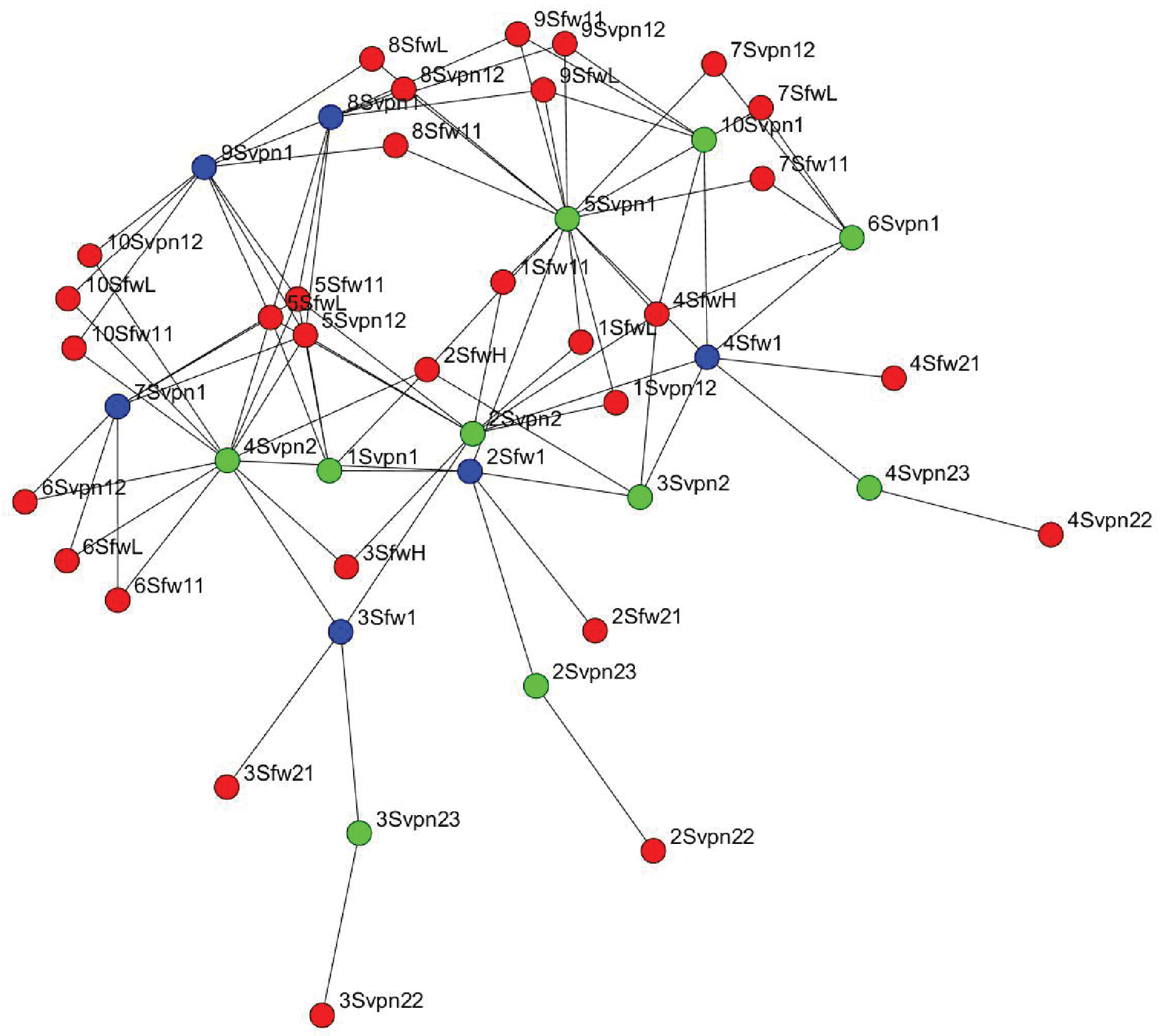}}
    \label{For IEEE-30 bus system}\\
   \caption{Different distributed coloring algorithm on diversity graph of IEEE-14 bus system}
  \label{fig1} 
  \vspace{-1em}
\end{figure*}

Then, we modeled our security graph $M$ 
for the IEEE-14 bus system based on the assumption proposed in section III-B. Next, we extracted the diversity graph $G$ from $M$  which was only based on the connectivity of SMs.

We have assumed steady-state probabilities for intrusion scenarios of $p_i$ to calculate the vulnerabilities. In Table III, we prioritize each $p_i$ from top-to-bottom order based on their security strength. As the security strength increases, it is less probable to attack. Hence, The attack likelihood $\pi$ is assumed in such fashion.

In this work, we had assumed that there exists a limited number of software packages(color) to diversify the SM and each of those assigned a certain integer value based on their security strength. The integer value assigned to available color for IEEE-14 bus system is, $c=\{Green, Blue, Red, Purple, Yellow \} \iff \{10, 8, 6 ,4, 2\}$

Finally, we apply the different distributed algorithm in $G$ and calculate the payoff (security index) of each SM. Figure $8(a)$, $8(b)$, $8(c)$ and $8(d)$ show diversity graph $G$ of IEEE-14 bus system after applied the randomized coloring algorithm, generic greedy coloring algorithm, sequential coloring algorithm and graph coloring game, respectively.

\begin{table}
\captionsetup{aboveskip=5pt, belowskip=5pt}
\caption{Impact factor calculation of IEEE 14-bus system} 
\begin{footnotesize}
  \begin{tabular}{|p{1.0cm}  p{1.5cm}  p{1.5cm}   p{1.2cm}   p{1.5cm}  |  } \hline
  \bf{Sub.}	 &  \bf{Associated Bus} &  \bf{LOL(MW)}	 &  \bf{$L^*$}  &  \bf{Impact factor($\gamma$)}   \\ \hline
\rowcolor{Gray}
1 &1& 0.5& 3.00 & 0.000071\\
2 &2& 5& 1 & 1.0 \\
\rowcolor{Gray}
3 & 3 & 94.24&3.059 & 0.2427 \\ 
4 &  4,7,8,9& 29.50& 1&1.0\\ 
\rowcolor{Gray}
5 &  5,6& 11.20&1.8 & 0.1050\\ 
6 &  10& 9.00& 3.066& 0.0019\\ 
\rowcolor{Gray}
7 &  11& 3.5& 3.062& 0.00027\\ 
8 &  12& 6.1& 3.04& 0.00092\\ 
\rowcolor{Gray}
9 &  13& 13.5& 3.059& 0.0044 \\ 
10 &  14&14.9& 3.066& 0.0053\\  \hline
  \end{tabular}
    \label{critical}
\end{footnotesize}
\vspace{-1em}
\end{table}

\begin{table}
\begin{footnotesize}
\captionsetup{aboveskip=5pt,belowskip=5pt}
\caption{Steady State Probabilities for Security Mechanism}
  \begin{tabular}{|p{2.5cm}  p{3.0cm}  p{2.1cm} |}\hline
  \bf{Attack start from} & \bf{Security Mechanism} & \bf { $\pi$}\\\hline
  \rowcolor{Gray}
SCADA firewall& $S_{fwH}$, $S_{fwL}$&0.1 \\
VPN&$S_{vpn1}$, $S_{vpn2}$ & 0.2\\
\rowcolor{Gray}
System firewall&$S_{fw11}$, $S_{fw21}$, $S_{fw1}$ & 0.5\\
System authentication&$S_{vpn22}$, $S_{vpn12}$, $S_{vpn23}$ & 0.8\\\hline
  \end{tabular}
  \label{critical}
\end{footnotesize}
\vspace{-1em}
\end{table}
\vspace{-.5cm}
\subsection{Scenario Analysis}
In this section, two possible scenarios had been analyzed in IEEE-14 bus system to show how the optimal diversity had been achieved from graph coloring game by: $(i)$ reducing the attack propagation and $(ii)$ increasing the security of the network.
\subsubsection{Reduce attack propagation}
Let us consider a scenario where the attacker wants to take control of the most critical substation $2$ within a limited capability, $k$. The attacker can access the SM whose strategy profile integer value $c$ is equal or less than $k$. Here, we assume the attacker capability, $k=8$. This scenario assumes substation connectivity incorporate with loss of load and mapping of color. The attacker is able to take control of substation $2$ by accessing either the SCADA firewall ($2S_{fwH}$), VPN ($2S_{vpn2}$) or system firewall ($2S_{fw1}$). After accessing $2S_{fwH}$  and $2S_{fw1}$, It is possible for an attacker to propagate the attack, as substation 2 is connected to substation $1, 3, 4$ and $5$. Also, the attacker needs to take control other substations first in order to access $2S_{vpn2}$.  

Table IV shows total loss of load of IEEE-14 bus system when an attacker gets access to substation 2 with his limited capabilities under different distributed coloring algorithm. In this table, column III represents which substations need to be compromised before accessing substation 2 and column IV represents which other substations had been affected by accessing substation 2. From fig. $8(d)$, it was observed that the color assigned by neighbor nodes of $2S_{fwH}$ and $2S_{fw1}$ are green whose security strength integer value is $c=10$. Hence, the attacker can take control the substation $2$ by accessing both SCADA firewall and system firewall but not able to propagate his attack into other substations due to his limited capabilities. But by using other traditional coloring algorithm, the attacker is able to access the substation 2 and propagate his attack to other substations. From table IV, it concluded that a graph coloring game reduces attack prorogation and minimizes loss of load by allocating appropriate software packages to the security mechanism.  
\begin{table}
\begin{footnotesize}
\captionsetup{aboveskip=5pt, belowskip=5pt}
\caption{ Total loss of load on accessing substation 2 of IEEE-14 bus system}
\setlength\arrayrulewidth{.9pt}
  \begin{tabular}{|p{1.8cm}  p{2.0cm} p{0.8cm} p{1.0cm}  p{1.2cm} |}\hline
  \bf{Distributed Algorithm} &\bf{Access security mechanism} &\bf{Attack start from} &  \bf{Attack propagate} &  \bf {Total $P_{lol}$ $(MW)$}   \\  \hline
\rowcolor{Gray}
Coloring game&$2S_{fwH}$, $2S_{fw1}$ & - &-& 5.0\\ 
Greedy&$2S_{fwH}$, $2S_{fw1}$, $2S_{vpn2}$  & 1,3,4,5&1,3,4,5& 140.44\\        
\rowcolor{Gray}
Sequential&$2S_{vpn2}$&1,5&-& 16.70\\
Random&$2S_{vpn2}$ &1,3,4,&-& 129.24\\ \hline
  \end{tabular}
  \label{critical}
\end{footnotesize}
\vspace{-1em}
\end{table}
\subsubsection{Increase the security} We analyze different scenarios of cyber-attack in single and multiple substation on IEEE-14 bus system to show how the diversity provided by the graph coloring game introduced difficulty for an attacker to access the entry point SMs of the substation. In all the scenarios, the attacker tries to get access of the SMs located on the entry point of the substation. Next, we calculated the security index of each SM for different distributed coloring algorithm by using Eq.2. According to Table V, for all the scenarios, the proposed graph coloring game allocates the most secure \textit{SM}s for protection against a cyber-attack. 
\begin{table}
\captionsetup{aboveskip=0pt, belowskip=5pt}
  \caption{Security index analysis of IEEE-14 bus system for different attack scenarios}
  \label{tbl:excel-table}
  \includegraphics[width=\linewidth]{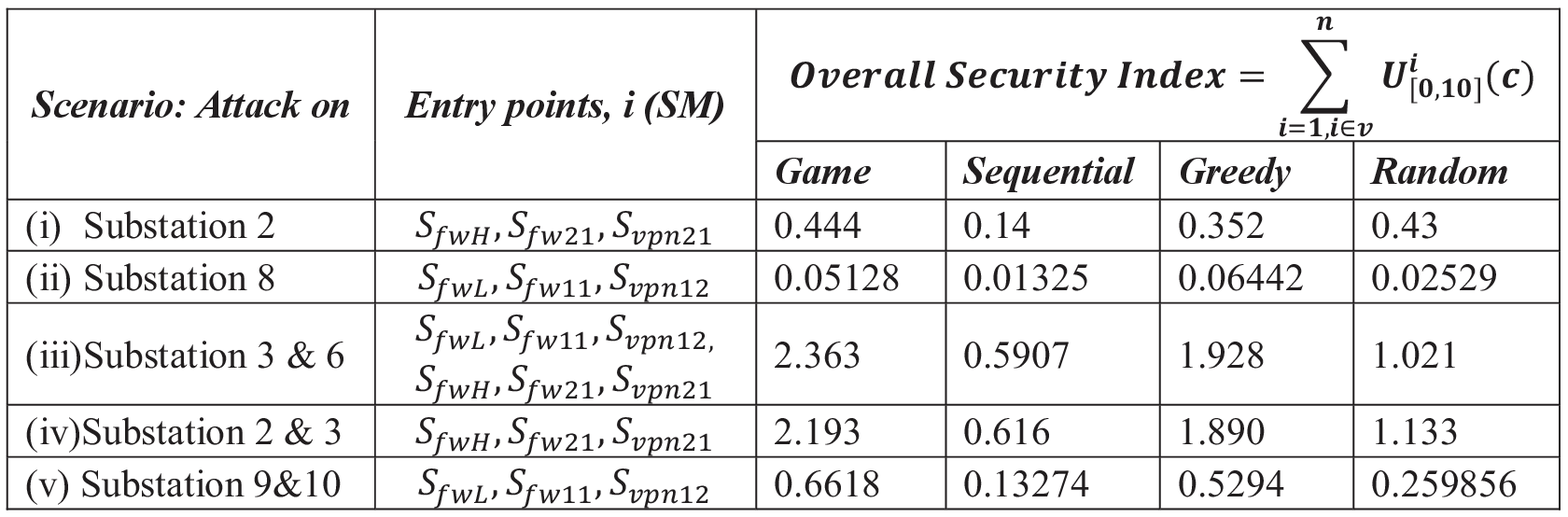}
  \vspace{-2em}
  \end{table}
\begin{table*}
\captionsetup{aboveskip=5pt, belowskip=5pt}
  \caption{Comparison of different distributed coloring algorithm for IEEE-118 bus system}
  \label{tbl:excel-table}
  \includegraphics[width=\linewidth]{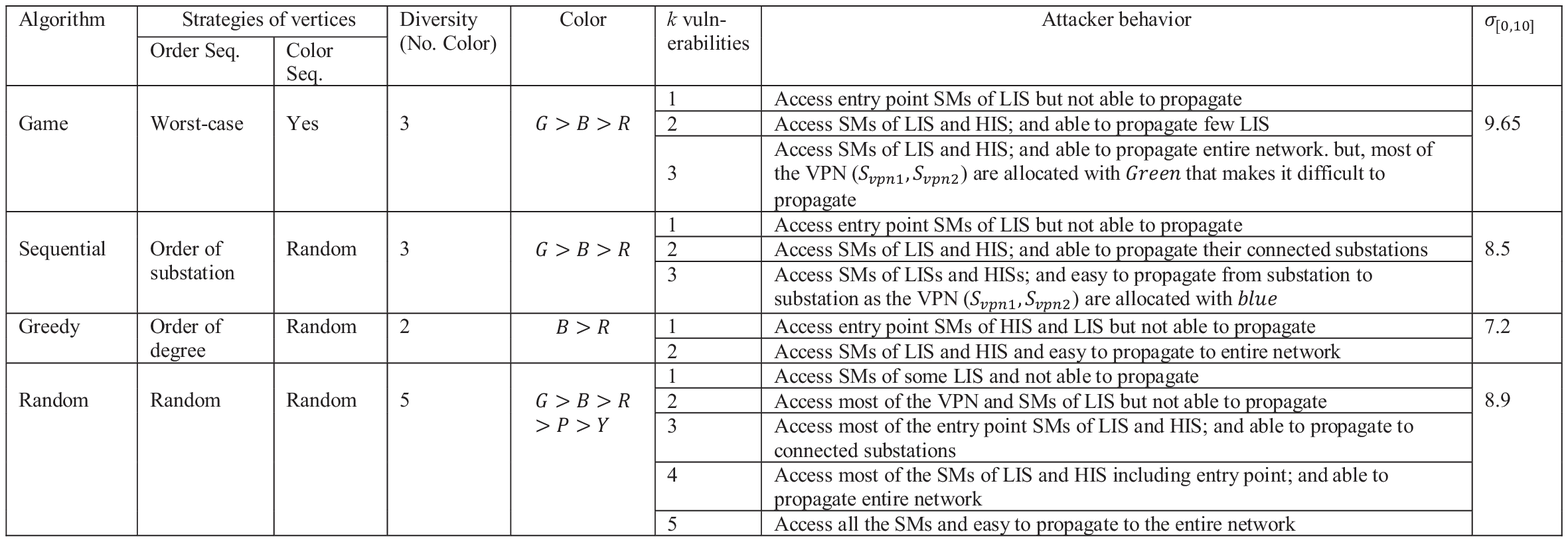}
  \vspace{-2em}
  \end{table*}
\vspace{-1em}
\subsection{Result analysis}
We have compared different distributed coloring algorithm by analyzing the attacker behavior against $k$ vulnerabilities on IEEE-118 bus system; and also by calculating the cumulative security index$(\sigma)$ for the entire diversity graph. The comparison of different distributed coloring algorithm is shown in Table VI. In this table, column $IV$ and column $V$ represents the number of color and which color required to diverse the entire graph, respectively. Column $VI$ represents the number of unique vulnerabilities. For example, in graph coloring game, $k=1$ describes an attacker able to access all the entry point SMs those are allocated with color red. 

From this table VI, for each algorithm, we had observed that when the maximum $k$ vulnerabilities is equal to diversity, then the attacker is able to take control the entire network by accessing all the SMs. Even though the diversity is same for the graph coloring game and the sequential algorithm, the diversity of SM in the graph coloring game makes the network more secure. This hinders the attacker capability to propagate the malware. 

The greedy coloring algorithm and the randomized coloring algorithm is able to diverse the entire network by using the least number and the most number of colors, respectively. But the cumulative security index$(\sigma)$ for greedy coloring algorithm is comparatively lower than other algorithm that implies the least secure allocation strategy of SMs.
\begin{figure}[ht!]
 \centering
  \includegraphics[width=\columnwidth,height=4.0cm]{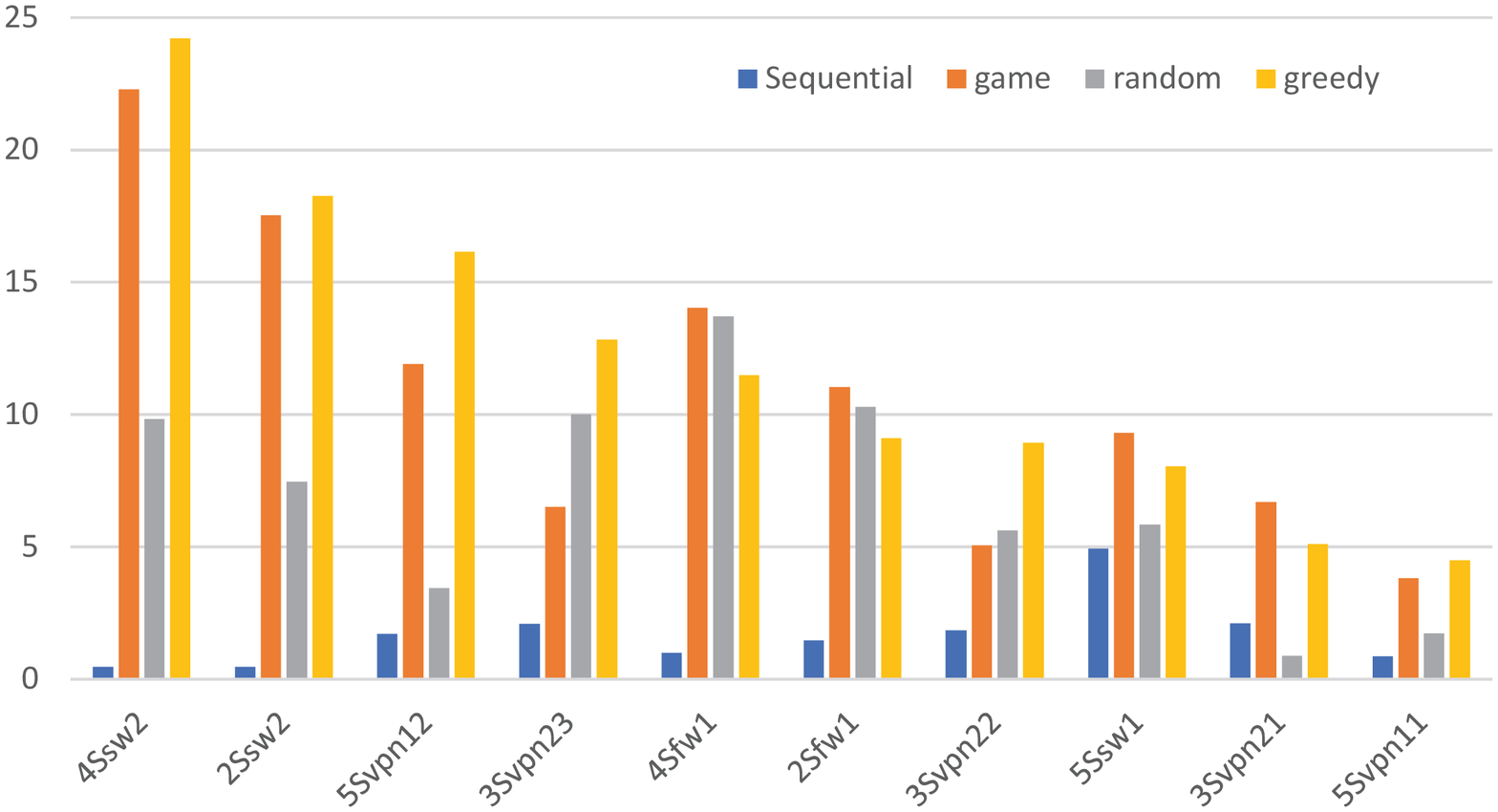}
  \caption{First ten highest security index $U^v$ \textit{SM} for different distributed algorithms} 
   \label{exp_fig}
   \vspace{-1em}
\end{figure}
For the graph coloring game, we observed that the cumulative security index is higher than all other distributed algorithms. 
Hence, this algorithm give the best possible software package allocation in each SM for IEEE-14 bus power grid network.

Figure 9 shows the first ten high security index \textit{SM}s of IEEE-14 bus system outputted by different distributed coloring algorithms. 
From this figure, we observed that most of the high-security index \textit{SM}s are located in HIS rather than LIS, If an attacker get access the HIS, he can cause more damage than accessing the LIS. Hence, the security index of \textit{SM} located in the HIS is higher by allocating more secure diverse \textit{SM}. According to the prioritization list, SCADA Firewall $(S_{fwH},S_{fwL})$ is more critical than VPN $(S_{vpn1},S_{vpn2})$. But according to security graph, if an attacker can access an VPN, he/she can also get access other substation which will cause most severe damage. Therefore, the VPN needs the most secure software combination to reduce the criticality of the entire network. From, figure 9, we also observed that the security index of the VPN located in substation 4 is the highest which indicates that the most secure software is allocated to this SM.

\section{Conclusion}
The security mechanism located within a ESP of an substation needs to be heterogeneous in order to increase the security of cyber assets in power grid network against a single shared software vulnerabilities. In this paper, we have applied different distributed coloring algorithms in our diversity graph to increase the effectiveness of SM heterogeneity. Among all the algorithms, the proposed graph coloring game provides the best diversity by increasing the security index and improving the attack tolerance of our power grid network. This security index can be used to minimize malware propagation and reduce loss of load, $P_{lol}$. 
In this analysis of the diversity problem, our model formulation is limited to defensive investment that leads to a additive level of expenditure by utilities. In future, we like to extend the study of diversity by introducing a new metrics that consider defensive investment too.

\end{document}